\documentclass[12pt,preprint]{aastex}
\usepackage{graphicx,natbib,amsfonts,amssymb,lscape}
\usepackage[figuresright]{rotating}
\usepackage{apjfonts}
\usepackage{color}
\usepackage{lscape}     
\usepackage{graphicx}
\usepackage{graphics}
\usepackage{psfig}
\usepackage{multicol}
\usepackage{natbib}

\newcommand{\ha}{\ifmmode H\alpha \else H$\alpha$\ \fi}
\newcommand{\hb}{\ifmmode H\beta \else H$\beta$\ \fi}

\newcommand{\hbb}{\ifmmode H\beta^{b} \else H$\beta^{b}$\ \fi}
\newcommand{\hbn}{\ifmmode H\beta^{n} \else H$\beta^{n}$\ \fi}
\newcommand{\ergs}{\ifmmode {\rm erg\ s}^{-1} \else erg s$^{-1}$\ \fi}

\newcommand{\kms}{\ifmmode {\rm km\ s}^{-1} \else km s$^{-1}$\ \fi}
\newcommand{\msun}{\ifmmode M_{\odot} \else $M_{\odot}$\ \fi}
\newcommand{\lv}{\ifmmode L_{\lambda}(5100\AA) \else $L_{\lambda}(5100\AA)$\ \fi}

\slugcomment{submitted to ApJ, on \today}

\usepackage{CJK}
\usepackage[T1]{fontenc}

\begin{document}
\begin{CJK}{UTF8}{gbsn}

\title{THE OPTICAL VARIABILITY OF SDSS QUASARS FROM MULTI-EPOCH SPECTROSCOPY. II. COLOR VARIATION }
\author{Hengxiao Guo (郭恒潇)$^{1,2}$, Minfeng Gu$^{1}$\\
$^{1}$Key Laboratory for Research in Galaxies and Cosmology, Shanghai Astronomical\\
         Observatory, Chinese Academy of Sciences, 80 Nandan Road Shanghai 200030, China \\
$^{2}$ University of Chinese Academy of Sciences, 19A Yuquanlu, Beijing 100049, China;
hxguo@shao.ac.cn, gumf@shao.ac.cn\\
} \shorttitle{The variability of SDSS quasars with multi-epoch spectroscopy} \shortauthors{Guo, et al.}

\begin{abstract}

We investigated the optical/ultraviolet (UV) color variations for a sample of 2169 quasars based on multi-epoch spectroscopy in the Sloan Digital Sky Survey (SDSS) data release seven (DR7) and data release nine (DR9). To correct the systematic difference between DR7 and DR9 due to the different instrumental setup, we produced a correction spectrum by using a sample of F-stars observed both in DR7 and DR9. The correction spectrum was then applied to quasars when comparing the spectra of DR7 with DR9. In each object, the color variation was explored by comparing the spectral index of the continuum power-law fit on the brightest spectrum with the faintest one, and also by the shape of their difference spectrum. In 1876 quasars with consistent color variations from two methods, we found that most sources (1755, $\sim 94\%$) show bluer-when-brighter (BWB) trend, and the redder-when-brighter (RWB) trend is only detected in 121 objects ($\sim 6\%$). The common BWB trend is supported by the bluer composite spectrum constructed from bright spectra than that from faint spectra, and also by the blue composite difference spectrum. The correction spectrum is proved to be highly reliable by comparing the composite spectrum from corrected DR9 and original DR7 spectra. Assuming that the optical/UV variability is triggered by fluctuations, RWB trend can likely be explained if the fluctuations occur firstly at outer disk region, and the inner disk region has not fully responded yet when the fluctuation being propagated inward. In contrast, the common BWB trend implies that the fluctuations are likely more often happening firstly in inner disk region.
\end{abstract}


\section{INTRODUCTION}
Quasars are the most luminous active galactic nuclei (AGNs) and they provide opportunity to explore far across the Universe. The study of variability is one of the most effective and important tool for revealing the nature of quasars \citep*[][and references therein]{ulrich1997}. The variability can happen at different timescales, from as short as hours usually occurring in jets of blazars, weeks for thermal changes in accretion disk, months for stochastic processes, to several years for global accretion rate changes or lens intersecting time \citep{rees1984,krolik1991,hawkins1996,kawaguchi1998,gupta2005,kelly2009}. However, the mechanism of variability is still inconclusive. By investigating the spectral variability of the optical/UV continuum, the changes in global accretion rate have been proposed to explain the optical/UV variability in accretion disk models \citep{pereyra2006,li2008,sakata2011,zuo2012,gu2013}. In lines with these models, \cite{gaskell2008} argued that the variability may propagate at close to the speed of light, rather than on viscous timescale. The variability can also be due to the hot spots \citep{abramowicz1991,zhang1991,pechacek2008,pechacek2013} or localized coronal flares \citep{galeev1979,merloni2001,czerny2004} in accretion disk caused by the enhancement of mass accretion or disk instabilities. Alternatively, it could be caused by the thermal fluctuations driven by an underlying stochastic process \citep{kelly2009,macleod2010}, which is related to general accepted magnetorotational instability \citep{balbus1991}, or large localized temperature fluctuations \citep{dexter2011,ruan2014,sun2014}. 

The color variations can give us clues on emission mechanisms, and have been extensively studied by using the large sky coverage survey, such as photometric and spectroscopic data from Sloan Digital Sky Survey (SDSS)  \citep{wilhite2005,gu2011a,gu2011b,meusinger2011,sakata2011,bian2012,zuo2012,guo2014,ruan2014}. There seems to be general consensus in the community that bluer-when-brighter (BWB) trend is common in radio-quiet quasars \citep{webb2000,wilhite2005,schmidt2012,sun2014}, and also in blazars \citep{gu2006,rani2010,gu2011a,zhang2015}. However, the redder-when-brighter (RWB) trend, firstly discovered in an optically violently variable quasar 3C 446 \citep{miller1981}, has been found in flat-spectrum radio quasars (FSRQs) \citep{gu2006,rani2010}. Interestingly, based on the study of two-epoch spectra of 312 radio-loud (RL) and 232 radio-quiet (RQ) quasars, \cite{bian2012} found that half of sources follow BWB trend, while the other half show RWB trend. Moreover, there is no obvious difference in color variation between their subsamples of RL and RQ quasars. In a sample of 9093 quasars in SDSS Stripe 82, \cite{schmidt2012} confirmed the BWB trend and suggested that the color variation is remarkably uniform and independent of redshift, luminosity and black hole mass. In addition, they claimed that relatively strong color variation were usually companied with fast, low amplitude and brightness variabilities. In a similar sample, \cite{sun2014} found that the color variation at shorter timescales was bluer than that at longer timescales, implying that RWB trend may more likely occur in long-timescale variabilities.

Although the color variations have been extensively studied, the fraction of RWB is still unclear with different results from different samples and database, not to say the mechanism of RWB \cite[e.g.,][]{bian2012,schmidt2012}. While most of works were based on massive photometric data \citep{schmidt2012,zuo2012}, there are few from spectroscopic data. The biggest spectroscopic sample only consist of about 600 quasars based on either early SDSS spectroscopy or spectra taken from multi-telescopes \citep{bian2012,ruan2014}. The SDSS spectra enable us to study the color variations for large samples and in broader wavelength coverage than photometry, and more importantly by avoiding the contaminations of strong emission lines in photometric filters. Moreover, the large database in SDSS enable us to study the color variation at high significance by carefully selecting the samples with high S/N spectra and minimizing the contamination effects (e.g., host galaxies).
Therefore, to further study the color variation, we compiled a large sample of 2169 quasars with multi-epoch spectroscopy from SDSS\footnote{\tt http://dr9.sdss3.org/bulkSpectra}. In Section 2, we describe the quasar sample. The spectroscopic data analysis, and the correction of the systematic difference between SDSS data release seven (DR7) and data release nine (DR9) due to the different instrumental setup are given in Section 3. We show the results and discussions in Sections 4 and 5, respectively. Finally, we draw our conclusions in Section 6. Throughout this paper, the spectral index $\alpha_{\lambda}$ is defined as $f_{\lambda}$ $\propto$ $\lambda^{{\alpha}_{\lambda}}$ with $f_{\lambda}$ being the flux density at wavelength $\lambda$.

\section{SAMPLE}
In SDSS, some plates were entirely or partially observed for multiple times.  This usually occurred when the signal-to-noise ratio (S/N) of the first epoch is not sufficiently reach the lower limit required by the survey, or it is inherently a part of survey plan \citep{paris2012,dawson2013,guo2014}. Consequently, many sources have multi-epoch spectroscopy in SDSS, which can be used to study quasar variability. In this paper, we combined SDSS DR7 and DR9 to enlarge the quasar sample with multi-epoch spectroscopy. We started from searching SDSS DR9 quasar catalog for the objects observed in DR7 before as indicated in the catalog \citep{paris2012}. By adding the quasars observed at least twice in SDSS DR7 quasar catalog \citep{shen2011}, however not included in DR9 quasar catalog, we constructed our parent sample of about 17000 quasars. All these sources have been observed at least twice in DR9 and/or DR7.

The further refinements on the sample were committed in several ways. Firstly, the sources with a non-zero $\rm ZWARNING$ flag or broad absorption lines were excluded due to the uncertain redshift determination and contamination of absorption lines \citep{reichard2003}. Secondly, only those spectra with high S/N $\geq$ 10 were selected to ensure the data quality. Thirdly, a redshift cutoff is set at $z\ge0.3$ to reduce the contamination of host galaxy. Instead of using the flux density at a fixed rest frame wavelength, we used the integrated flux density of overall spectrum to study the source variability. We define the integrated flux density for the studied spectrum as $f_{\rm int}$ = $\Sigma^{\rm n}_{\rm i=1}~ f_{\lambda_{\rm i}}$, in which $f_{\lambda_{\rm i}}$ is the flux density at wavelength $\lambda_{\rm i}$, and n is the total number of wavelength points. To increase the significance of variability in individual quasars, we only consider the spectra with the brightest and faintest integrated flux density. The variability between these two epochs is defined as $\Delta f$ = $(f_{\rm int,b}-f_{\rm int,f})/f_{\rm int,f}$, where $f_{\rm int,b}$ and $f_{\rm int,f}$ are integrated flux density at the brightest and faintest epochs, respectively. To ensure the reliability of variations, we selected those sources with $\Delta f > 10\%$. Moreover, twenty objects were excluded due to the significantly convex spectral shape, usually at the faintest epoch. The convex shape is difficult to fit with a single power-law. These features are possibly caused by dust extinction, and a systematic study on these sources will be present in a forthcoming paper (Guo et al. 2015, in preparation). Our final sample consists of 2169 quasars, of which 789 objects have both the bright and faint spectra in DR7, while 1380 sources have one epoch in DR7 and the other in DR9. The redshift of our sample sources ranges from 0.3 to 4.1. In our sample, 202 quasars ($\sim 9.3\%$) are radio loud according to the radio loudness $R=f_{\rm 6 cm}/f_{\rm 2500 {\AA}} \ge 10$ obtained directly from \cite{shen2011}, with $f_{\rm 6 cm}$ and $f_{\rm 2500 {\AA}}$ being the flux density at rest-frame 6 cm and 2500 \AA, respectively.

\section{RECALIBRATION ON BOSS SPECTRA AND DATA ANALYSIS}

DR9 quasar catalog is the first quasar catalog of the Baryon Oscillation Spectroscopic Survey \cite[BOSS,][]{dawson2013}. Compared to SDSS DR7, there are two main modifications in BOSS\footnote{\tt https://www.sdss3.org/dr9/whatsnew.php}: (1) original SDSS spectrographs were replaced by BOSS spectrographs in 2009, which are fed by 1000 fibers of 2" entrance aperture, rather than 640 fibers of 3" aperture. The CCDs were substituted with new devices with higher throughput and smaller pixels. The gratings were changed into volume-phase holographic gratings; (2) new target selection algorithms with a larger range of luminosities and colors were used, which mainly focus on high redshift objects to measure the baryon acoustic oscillation.

\begin{figure}
\centering
\includegraphics[]{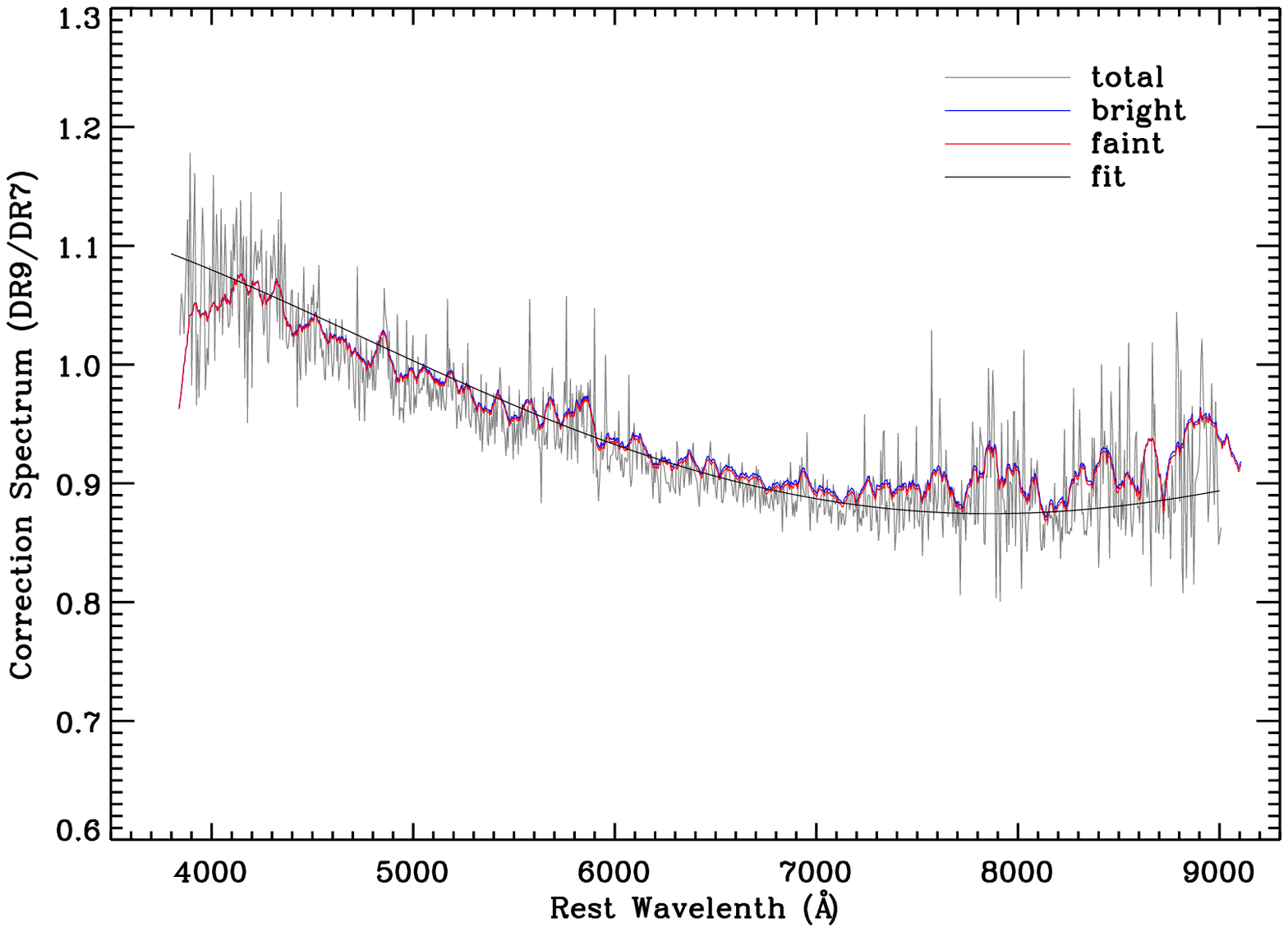}\\
\caption{The correction spectrum between DR7 and DR9 derived from 80 F-type stars (grey line). The smooth solid black line represents the  four-order polynomial fit, i.e. Equation (1), which was used to correct the systematic difference between DR7 and DR9. The blue and red lines represent the correction spectra constructed with 19 bright and 61 faint F-stars, respectively (see text for details), which are smoothed every 50 points to help distinguish from the overall correction spectrum.   
\label{fig:correct80}}
\end{figure}

Different devices and algorithms in BOSS lead to substantial differences in calibration from that of SDSS-DR7 targets. Two effects are crucial for the calibration: (1) excess flux in the blueward of BOSS spectrum; (2) the application of washers and the focal-plane offsets in plate position for the BOSS quasar fibers, which accounts for a warping of the quasar target spectrophotometry relative to the standard stars \citep{paris2012,dawson2013,margala2015}. Both effects will lead to bluer spectra. The excess light is about 10$\%$ at observed wavelength 3600 {\AA}, and decreases with the increasing wavelength in BOSS spectra \citep{paris2012}. The fiber offsets are originally designed to improve throughput in the Lyman-$\alpha$ forest and reduce the atmospheric differential refraction for quasar targets, however, these offsets are not applied to the standard stars \citep{dawson2013}. Moreover, the current pipeline for DR9 (idlspec2d v5$\_$4$\_$45) have not taken this into account. Such systematic flux difference between BOSS and DR7 are therefore nontrivial and must be corrected in order to study the quasar variability when comparing DR9 with DR7 spectra. Instead of using standard stars, we selected a sample of about 30,000 objects with a $\rm QSO\_LIKE$ flag from DR9, which were quasar candidates based on the target selection criteria of BOSS \citep{paris2012}, however finally proved to be stars (usually called "failed quasar"). These sources were then cross-matched in DR7. We found that 400 stars were previously observed in DR7, of which 80 objects are F-type stars. The systematic correction spectrum was generated by averaging the correction spectrum of individual F-stars, which is the flux ratio of BOSS to DR7 spectra (see Fig. \ref{fig:correct80}). The reason to use only F-type stars is that they are usually blue, stable, observed at high S/N and selected from high galactic latitude. Moreover, they are often used in the calibration procedure of SDSS. In fact, we find that the systematic correction spectrum generated from 80 F-type stars is very similar to that from all 400 stars, which includes various types of stars. 

As shown in Fig. \ref{fig:correct80}, the systematic correction spectrum is fitted with a fourth-order polynomial 

\begin{equation}
\rm f_{c,\lambda} = a+b\lambda+c\lambda^2+d\lambda^3+e\lambda^4,
\end{equation}
 
where $\rm f_{c,\lambda}$ is the average flux ratio at observed wavelength $\lambda$, a = 0.81, b = 3.40 $\times 10^{-4}$, c = $-~ 1.12 \times 10^{-7}$, d = 1.23 $\times 10^{-11}$ and e = $- ~4.38 \times 10^{-16}$. This however can only be applied in 3800 - 9200 {\AA}, since the wavelength coverage are 3800 - 9200 {\AA} and 3600 - 10500 {\AA} for DR7 and BOSS, respectively. Our systematic correction spectrum indeed shows that BOSS spectra are systematically bluer at short wavelength, consistent with the differences in the composite spectra generated for DR7 and DR9 quasars \cite[see Fig. 5 in][]{paris2012}. 
Since the correction spectrum is essential in our study, it needs to be carefully checked whether it depends on source luminosity. By separating the F stars to two groups with luminosity larger (19 stars) or smaller (61 stars) than the median value, we found that the correction spectra of two groups are in good consistence (see Fig. 1). This strongly indicates that our correction spectrum is independent of source luminosity.

\begin{figure}
\centering
\includegraphics[]{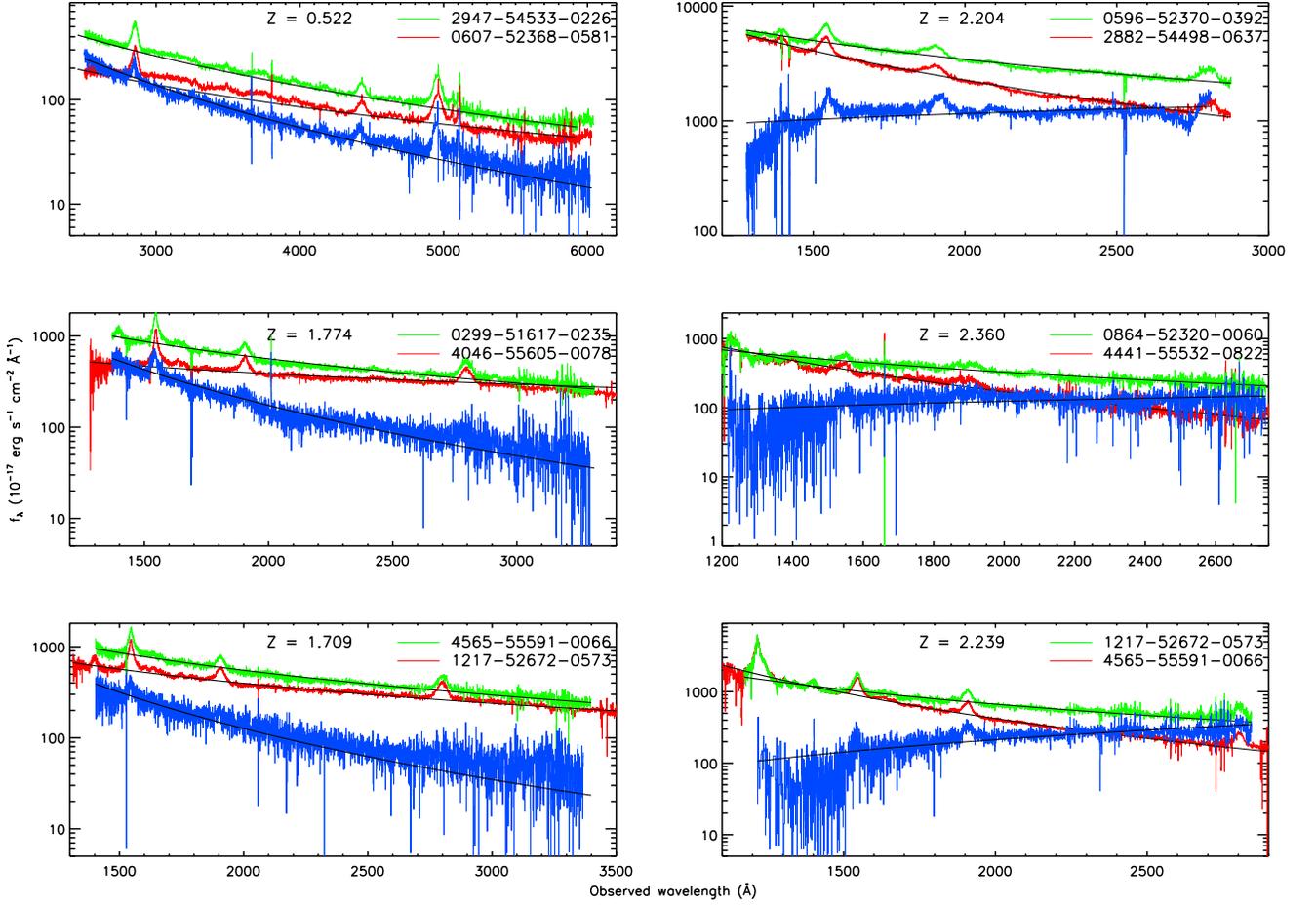}\\
\caption{
Examples of color variations. The left column shows BWB trend in three objects, while the right one is for three RWB sources. In each panel, the red and green curves are spectra at two epochs, and the blue one is their difference spectrum. The optical/UV Fe II emission lines and Balmer continuum were subtracted from the spectra. The black lines are power-law fits on the continuum.
\label{fig:example}}
\end{figure}

In this paper, we focus on the variation of continuum shape when flux varies. Therefore, only the continuum were fitted, and the measurements on the emission lines were not performed. When BOSS spectra are involved, we firstly corrected the systematic difference between BOSS and DR7 by dividing BOSS spectrum with the systematic correction spectrum, i.e. Equation (1). Then, all the spectra were transferred into source rest frame after correction for Galactic extinction using the extinction map \citep{schlegel1998} and reddening law \citep{cardelli1989}. Several line-free windows were selected to obtain a pseudo-continua, then fitted with a power-law continuum, UV/optical Fe II emission and a Balmer continuum together \cite[see also][]{chen2009,guo2014}. The $\chi^2$ minimization was applied to obtain the best fit by using the IDL package $\emph mpfitexpr$\footnote{\tt Perform Levenberg-Marquardt least-squares fit to arbitrary expression}. The spectral index $\alpha_{\lambda}$ of the power-law continuum can be obtained from the best fit, as shown in Fig. \ref{fig:example} for examples.

\section{RESULTS}
\subsection{Color Variation}

\begin{figure}
\centering
\includegraphics[]{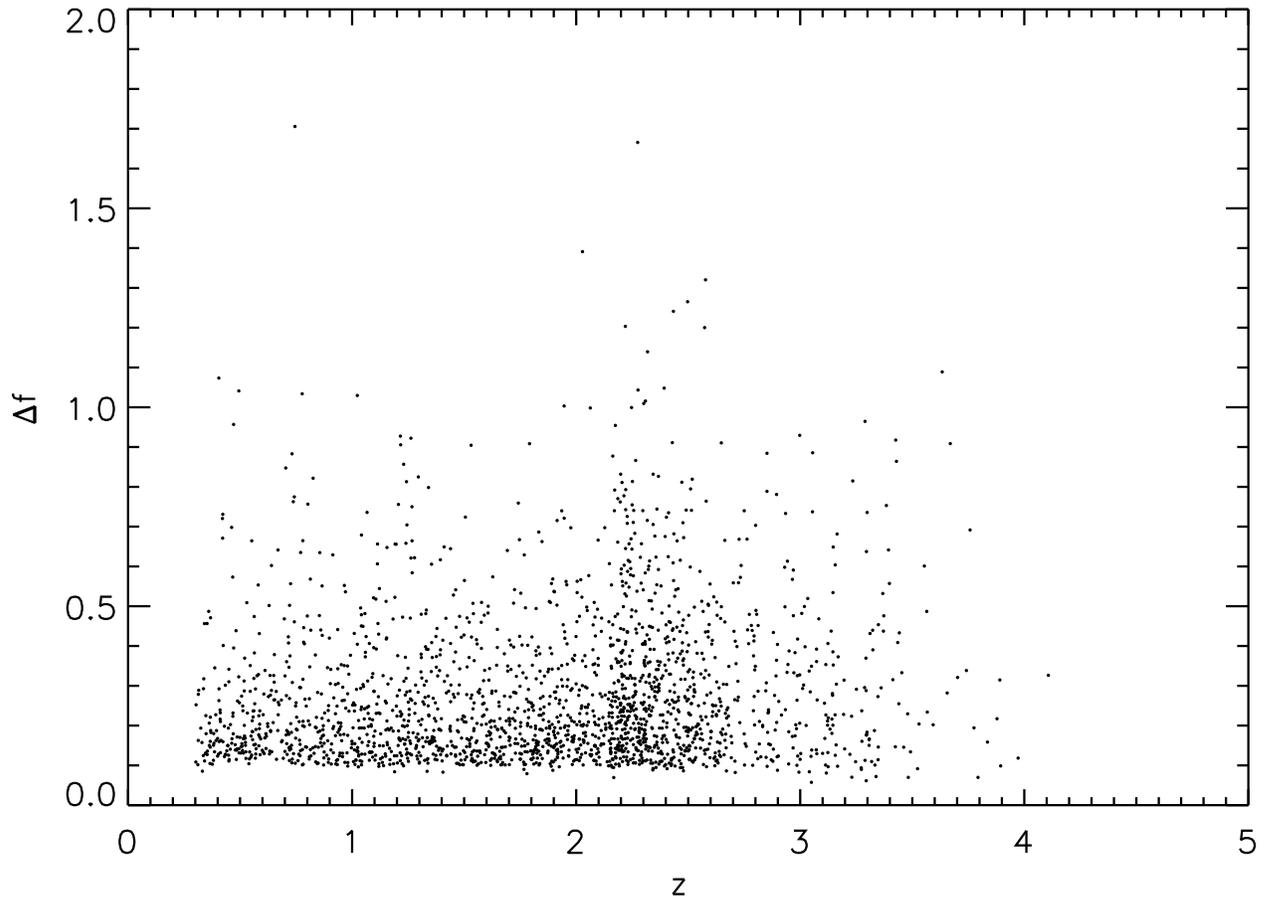}\\
\caption{
Variability versus redshift. As designed, our sources are limited to have redshift $z\ge0.3$ and variability $\Delta f>10\%$ (see text for details). Most sources at $z>2.15$ are from BOSS.
\label{fig:zdistr}}
\end{figure}

The distribution of the variability $\Delta f$ with redshift is shown in Fig. \ref{fig:zdistr}. The variability ranges from 10\% to 170\% with most sources below 50\%. Most objects with redshift higher than 2.15 are from BOSS as expected, since it was designed to detect Lyman-$\alpha$ forest in high redshift quasars.

\begin{figure}
\centering
\includegraphics[]{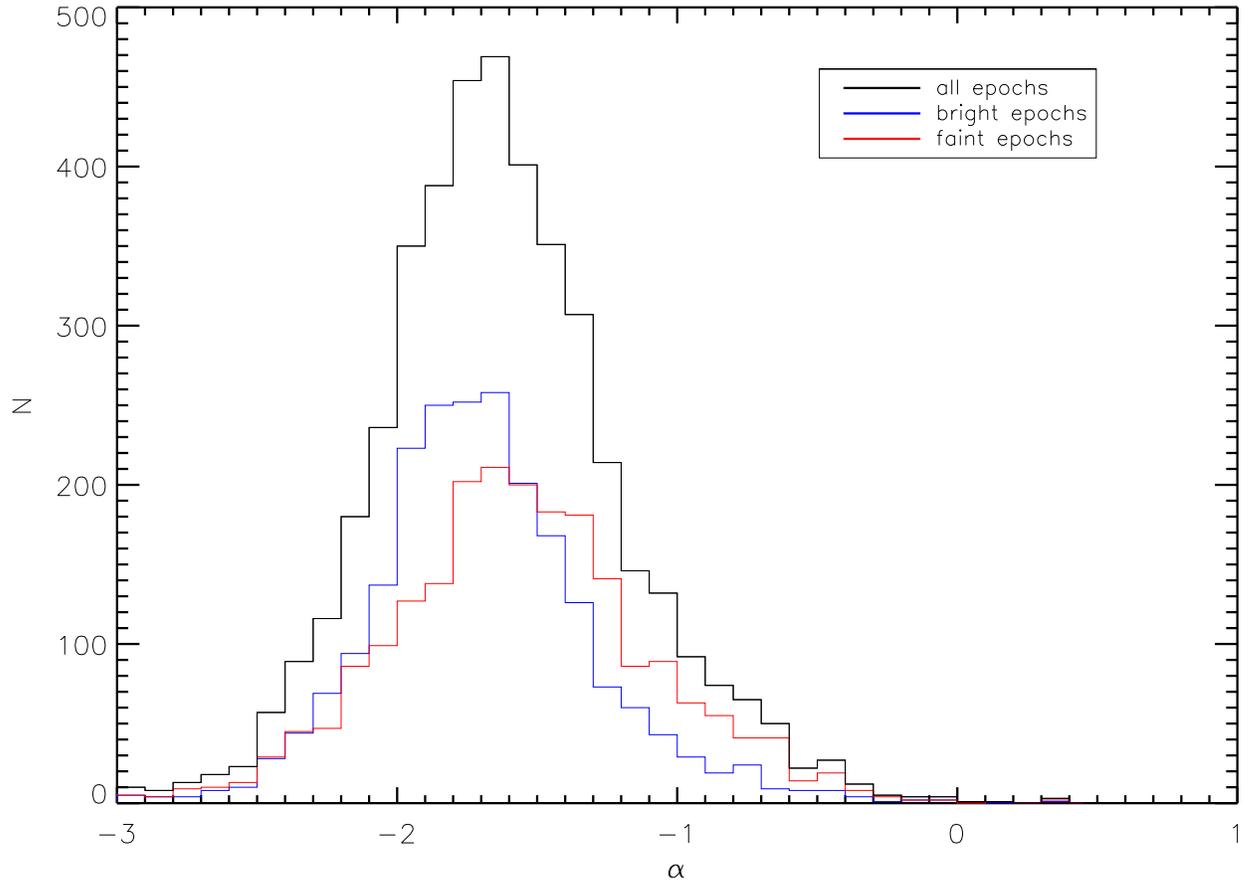}\\
\caption{The histogram of spectral index of continuum for our sample. The black solid line is for all epochs, while the blue and red solid lines are for bight and faint epochs, respectively.
\label{fig:distr}}
\end{figure}

The color variation in individual quasars was firstly determined from the difference of the spectral index between the bright and faint epochs $\Delta \alpha = \alpha_{\rm b} - \alpha_{\rm f}$, in which $\alpha_{\rm b}$ and $\alpha_{\rm f}$ are spectral index at bright and faint epochs, respectively. We found that 1782 quasars show BWB trend ($\Delta \alpha < 0$), while the rest 387 objects exhibit RWB trend ($\Delta \alpha > 0$). The color variations were further checked by the power-law fit to the difference spectrum between bright and faint epochs (i.e., bright minus faint spectra, see Fig. \ref{fig:example}). The spectral index of the difference spectrum $\alpha_{\rm d}$ show that 1755 quasars are confirmed to have BWB trend ($\Delta \alpha < 0$, and $\alpha_{\rm d} < 0$), and 121 objects with RWB trend ($\Delta \alpha > 0$, and $\alpha_{\rm d} > 0$, see Figs. 2 and 5). By only considering the quasars with consistent color variations from two methods, we found that majority of sources (1755/1876, $\sim 94\%$) show BWB trend, consistent with previous results based on SDSS \cite[e.g., ][]{zuo2012,ruan2014}. This result is also supported by the bluer spectra at bright epoch than at faint epoch, with mean values of the spectral index, $-1.72$ and $-1.55$ for the former and the later, respectively (see Fig. \ref{fig:distr}).

\subsection{Composite Spectra}

To further study the color variations in our sample, we constructed composite spectra separately for bright and faint epochs. In this work, we used the geometric mean spectrum in order to preserve the global continuum shape, instead of the arithmetic mean spectrum, which preserves the relative fluxes of emission features. The geometric mean spectrum was generated following the procedure in \cite{vanden2001}, including rebinning the individual spectra to source rest frame, scaling the spectra, and finally stacking the spectra into the composite with $<f_{\lambda}> = (\prod_{i=1}^{n}f_{\lambda ,i})^{1/n}$, where $f_{\lambda, \rm i}$ is the flux of each spectrum at wavelength ${\lambda}$ and n is the total number of spectra in spectral bins (see Fig. \ref{fig:diff}). The composite difference spectrum was also derived in similar way. The composite spectra for both bight and faint epochs were firstly scaled at 2250 \AA. However, to distinguish two composites, we multiple a scalar of 0.76 on the faint epoch, which is the mean value of the flux ratio of bright to faint spectra at 2250 {\AA} (see Fig. \ref{fig:diff}). 

We fitted composite spectra with a power-law on several line-free regions, which were selected as $1350 - 1370$ {\AA}, $1455-1470$ {\AA}, $1680 - 1720$ {\AA}, $2160-2180$ {\AA}, $2225-2250$ {\AA}, $4000-4050$ {\AA} and $4210-4230$ {\AA}. The region bluer than $\rm Ly{\alpha}$ 1216{\AA} was not used due to the contamination of the absorption features. Moreover, the wavelength longer than $\rm H{\beta}$ was also not included, since it might be better fitted by a different power-law perhaps due to the host galaxy \citep{vanden2001}. As shown in Fig. \ref{fig:diff}, the bright composite spectrum is steeper than the faint one with the spectral index of $\alpha_{\rm c,b} = -1.72 \pm 0.07$ and $\alpha _{\rm c,f} = -1.54 \pm 0.03$, respectively. This strongly supports our finding that most of our sources show BWB trend. While the faint composite is consistent with the composite spectrum in \cite{vanden2001} ($\alpha_{\rm \lambda} = -1.56$), the bright one is bluer. We found that the composite difference spectrum is much steeper than any composites with a spectral index of $\alpha_{\rm c,diff} = -2.01 \pm 0.02$, consistent with previous results \cite[e.g.,][]{wilhite2005,ruan2014}. Its blue color gives additional support of general BWB trend in our sample. 

\section{Discussion}
\subsection{Correction spectrum}

DR9 spectra have been used for more than half sources in our sample, therefore, the reliability of the correction spectrum is crucial in our study. It can be checked by comparing the composite spectra at faint epoch constructed from BOSS-only and DR7-only spectra.  As shown in Fig. \ref{fig:all_dr7}, the composite spectrum from original BOSS spectra is obviously bluer than that of DR7 spectra, consistent with results of \cite{paris2012} (see their Fig. 5). Interestingly, the composite spectrum based on the corrected BOSS spectra is nearly identical to that of DR7. This strongly proves that the correction spectrum is highly reliable. It can be seen from Fig. \ref{fig:correct80} that the blue excess is about $10\%$ at 3800 {\AA}, however the flux could be underestimated at $\sim10\%$ at 9000 {\AA}, which is qualitatively consistent with recent analysis \citep{margala2015}. 

To further check the reliability of correction spectrum, we constructed the composite difference spectrum only for quasars with both bright and faint epochs in DR7, and compared with that of all sources. We found that two spectra are very similar (see Fig. \ref{fig:diff}), again indicating that the correction spectrum is highly reliable. Both spectra show significant variations in emission lines both for broad and narrow components (e.g. narrow Mg II and [O III] 5007 \AA ~ lines). The stronger line variations compared to \cite{wilhite2005}, are likely due to our larger sample size, and that we maximized the variability by selecting the brightest and faintest epochs. The correction spectrum is essential to correct the calibration difference between SDSS-DR7 and BOSS, even for the most recent data release 12, in which the calibration difference still remains \cite[see][]{alam2015}. It could be widely applied in future work to correct the systematic difference, especially in variability study.

\subsection{Uncertain color variation}

\begin{figure}
\centering
\includegraphics[]{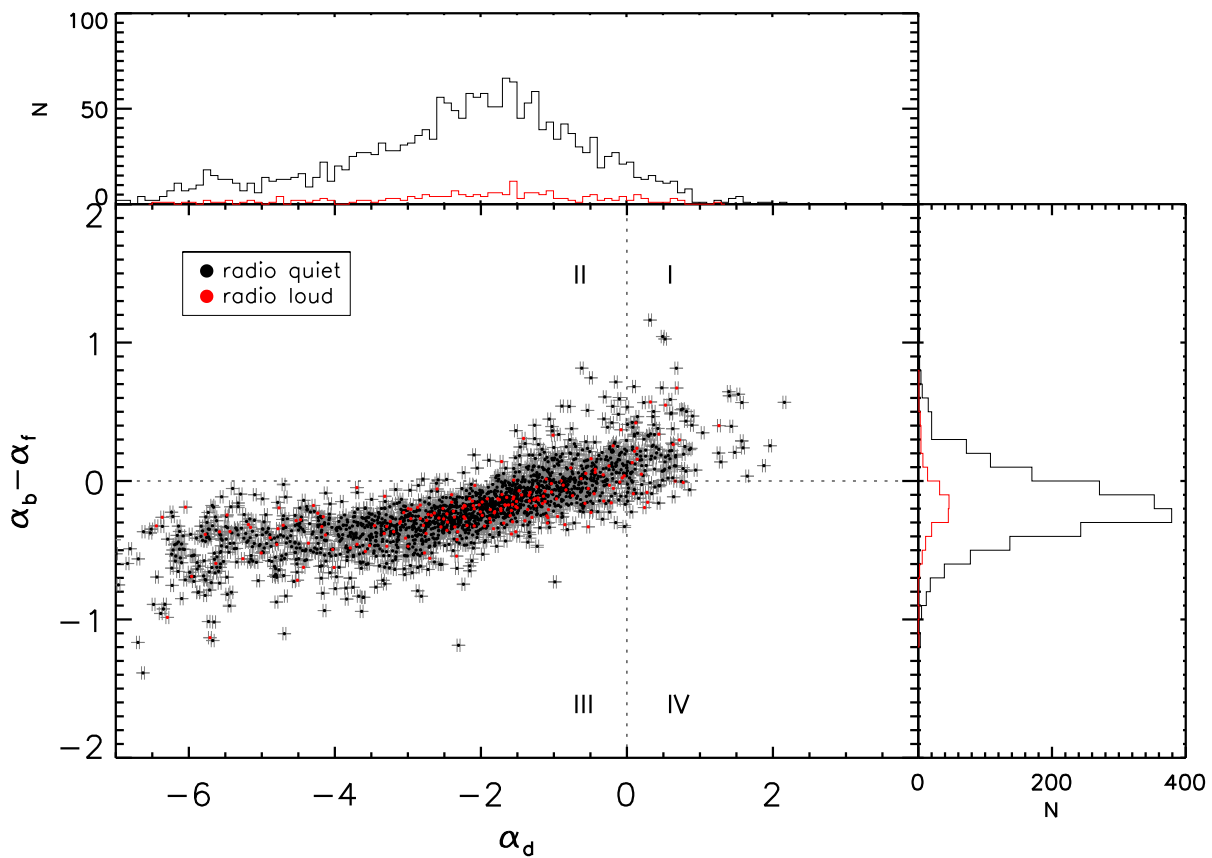}\\
\caption{Color variations in two methods. In main panel, the black and red data points are for RQ and RL quasars, respectively. The uncertainties are shown with grey error bars. The sources in region I and III have consistent color variations from two methods, thus were considered in our analysis. In contrast, the color variations of objects in region II and IV are uncertain due to various reason (see text for details), thus were excluded in analysis. Two side panels show the distribution of $\alpha_{\rm b} - \alpha_{\rm f}$ and $\alpha_{\rm d}$, from which it can be clearly seen that most sources have BWB trend, and there is no significant difference between RQ and RL quasars.
\label{fig:comp}}
\end{figure}

\begin{figure}
\centering
\includegraphics[]{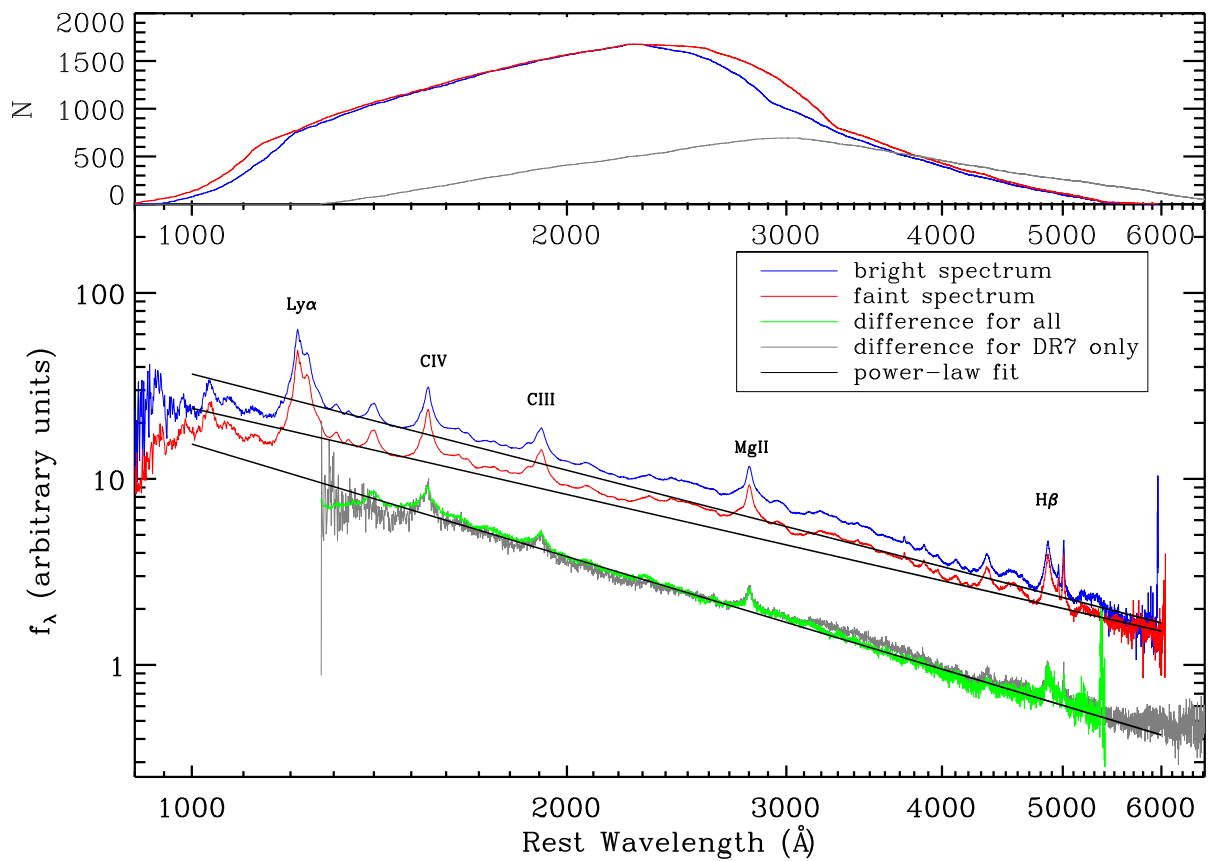}\\
\caption{Geometric mean composite spectra at bright and faint epochs, represented in blue and red colors, respectively. The grey and green curves are composite difference spectra for all sources, and for sources with both bright and faint epochs in DR7, respectively. All composite spectra were normalized at 2250{\AA}, however scalars of 0.76 and 0.30 are multiplied on the composite spectrum of faint epoch (red) and the composite difference spectrum (grey and green), respectively, to clearly separate from the one of bright epoch. The black solid lines are power-law fits on the continuum. The upper small panel shows the number of objects in each wavelength bins (1 \AA) when constructing the corresponding composite spectra.
\label{fig:diff}}
\end{figure}

\begin{figure}
\centering
\includegraphics[]{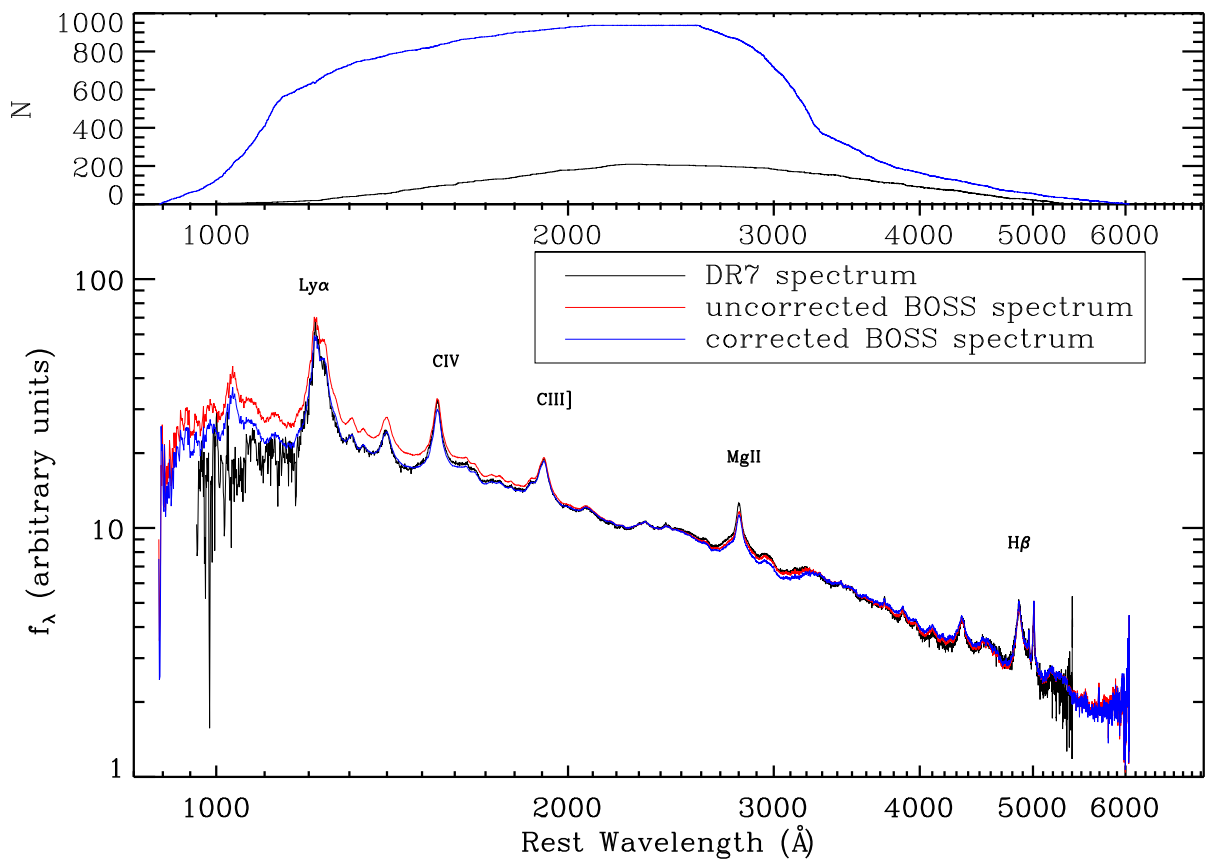}\\
\caption{Reliability of correction spectrum investigated by comparing the composite spectrum of DR7 and BOSS spectra. While the black line shows the composite spectrum from DR7 spectra at faint epoch, the red and blue ones are composite spectra at faint epoch for uncorrected and corrected BOSS spectra, respectively. All three composite spectra were normalized at 2250{\AA}. The good match between blue and black spectra strongly favors the reliability of correction spectrum. The upper small panel shows the number of objects in each wavelength bins (1 \AA) when constructing the corresponding composite spectra.
\label{fig:all_dr7}}
\end{figure}

As shown in Fig. \ref{fig:comp}, the color variations are uncertain in 293 quasars with either $\Delta \alpha < 0$ and $\alpha_{\rm d} > 0$, or $\Delta \alpha > 0$ and $\alpha_{\rm d} < 0$. This may be caused by several factors. The power-law fit on quasar continuum strongly depends on the selected line-free windows, which however are contaminated by optical/UV Fe II (mainly in 2200 $-$ 3800 {\AA} and 4400 $-$ 5500 {\AA}), and Balmer continuum ($< 4000$ {\AA}). Although Fe II and Balmer continuum were also added in the overall fitting, the uncertainties in the spectral index will result in a large uncertainty in the color variation $\Delta \alpha$. In contrast, most of the Fe II, Balmer continuum, and emission lines are removed in difference spectrum of individual quasars. This will largely reduce the uncertainty in $\alpha_{\rm d}$. Moreover, although we performed single power-law fit on the continuum, the better fit can be obtained with a broken power-law in many cases, such as a flattening at 3000 {\AA} due to small blue bump \citep{vanden2004}. This will also result in a large $\Delta \alpha$ uncertainty, however has little influence on the general trend of difference spectrum. In addition, the uncertain color variations will happen when one spectrum intersects with the other, since the color variation wholly depends on the recognition of bright and faint spectra, which is defined from the integrated flux density. One spectrum can be regarded as bright spectrum because of its higher integrated flux although the flux at fixed wavelength is smaller than the other spectrum. For these reasons, we required consistent color variations for individual objects from two methods. Although this will reduce the source number, it can give more reliable results on the color variations.

\subsection{Comparison between BWB and RWB quasars}

\begin{figure}
\centering
\includegraphics[]{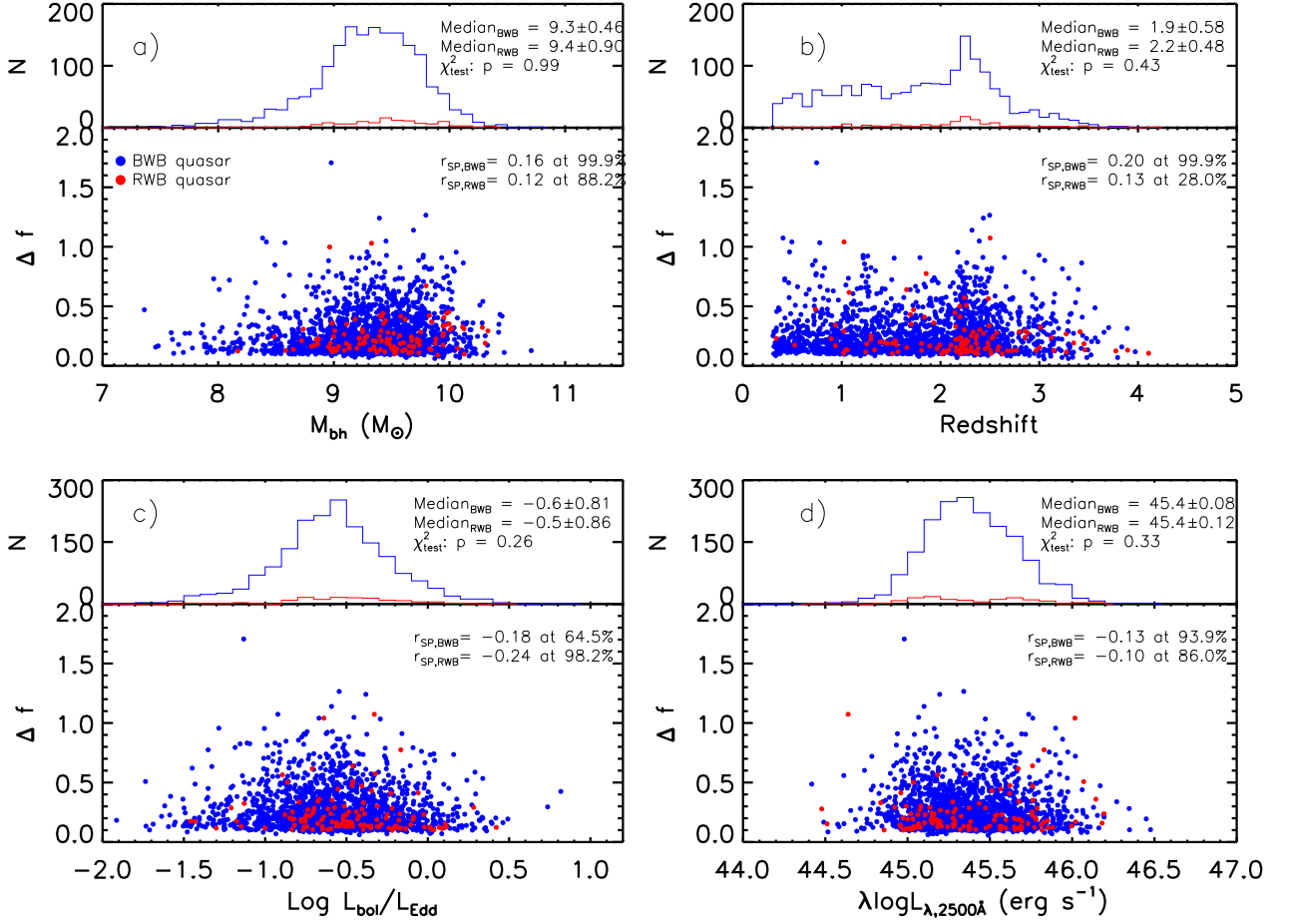}\\
\caption{ Variability versus: (a) black hole mass; (b) redshift; (c) Eddington ratio; (d) continuum luminosity at 2500{\AA}. The blue dots and lines are for BWB quasars, while the red dots and lines are for RWB ones. 
The results of Spearman correlation analysis (correlation coefficient and confidence level), are shown in each panel for BWB and RWB objects, respectively. In side-panels, the histogram of four parameters are shown for both BWB and RWB quasars, along with their median values. The average probabilities $p$ of 100-time $\chi^2$ test from comparing the randomly selected BWB samples with RWB quasars are also given.   
\label{fig:four}}
\end{figure}

The physical properties of RWB quasars are compared with those of BWB objects in Fig. \ref{fig:four}, in which the variability $\Delta f$ is plotted for 1755 BWB and 121 RWB quasars with black hole mass, redshift, Eddington ratio and continuum luminosity at 2500{\AA}, respectively. We find positive correlations in $\Delta f - M_{\rm bh}$ and $\Delta f - z$ both for BWB and RWB sources (see Fig. \ref{fig:four}a,b), while negative correlations in $\Delta f - L_{\rm bol}/L_{\rm Edd} $ and $\Delta f - L_{\rm 2500\AA}$ are present for both subsamples (see Fig. \ref{fig:four}c,d). These correlations are consistent with previous results \citep{guo2014,meusinger2013,vanden2004,wold2007}. For each parameters, RWB and BWB quasars have similar median values, indicating no significant differences in these two populations (see Fig. \ref{fig:four}). This is further studied using $\chi^2$ test assuming all quasars come from the same distribution. From RWB quasars, we randomly selected samples with same size as BWB sample for 100 times. Each randomly selected sample was compared with BWB quasars using $\chi^2$ test. The averaged probabilities from 100-time $\chi^2$ test ($p$, see Fig. \ref{fig:four}) indicate that there is no reason to reject the proposed hypothesis at the 0.05 significance level. This strongly favors similar distributions of four parameters in two populations. Therefore, the color variation behaviors are generally uniform in our sample, and independent of  black hole mass, redshift, Eddington ratio and luminosity.

\subsection{RWB trend}
The common BWB trend is consistent with previous results based on either photometric or spectroscopic data \cite[e.g., ][]{zuo2012,ruan2014}. There are several scenarios related with color variations. The spectral change may occur if two components with different spectra and variability timescale make up the continuum, for example, the relative stable small blue bump and the more variable primary continuum \citep{ulrich1997}. \cite{li2008} proposed that the change of accretion rate in accretion disk model can be responsible for the observed optical/UV variability, including the correlation of optical/UV variability amplitude with rest-frame wavelength, which is tightly related with BWB trend.  Whereas\cite{kelly2009} suggested that the variability is caused by thermal fluctuations driven by a stochastic process in accretion disk, e.g., a turbulent magnetic field, and their thermal timescale is about 10 $\sim$ $10^4$ days. Recently, a time-dependent inhomogeneous disk model with large temperature fluctuations was constructed, which can naturally explain the color variation and power-law composite difference spectrum \citep{dexter2011,ruan2014}. However, \cite{kokubo2015} argued that the intrinsic scatter of magnitude - magnitude plots for Stripe 82 quasars light curves is too smaller than the simulated scatter, and their inhomogeneous model cannot explain the tight inter-band correlation.

\begin{figure}
\centering
\includegraphics[]{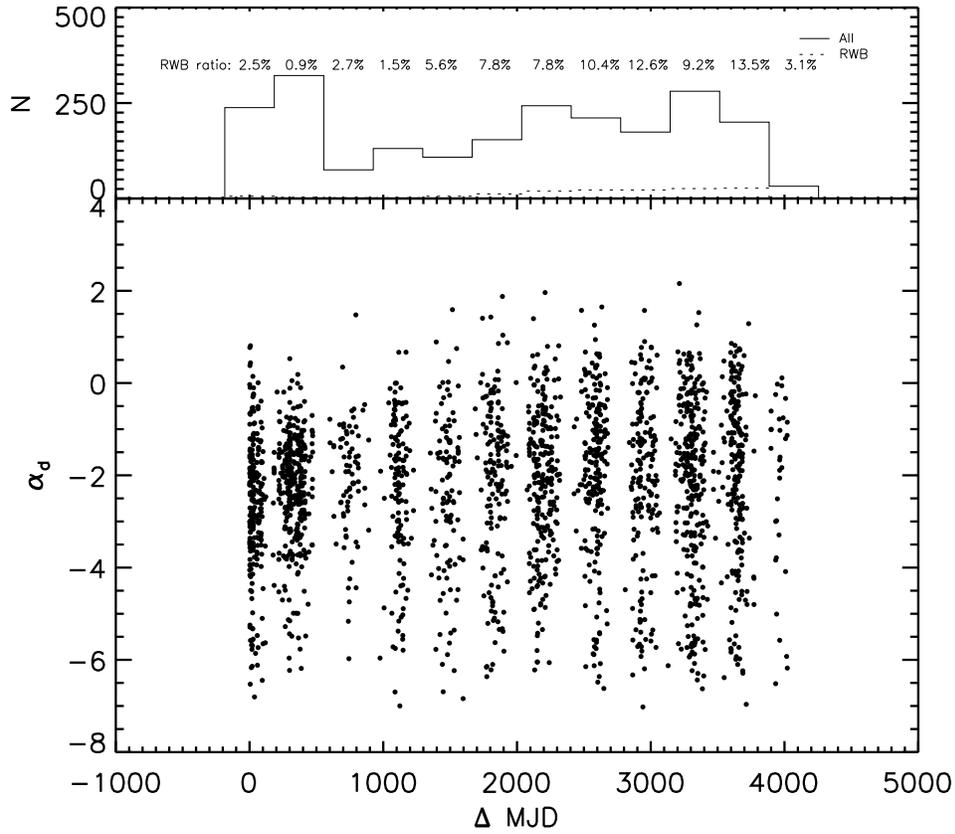}\\
\caption{ Correlation between the spectral index of difference spectrum and time separation. The side-panel shows the histogram of all quasars (solid line) and RWB objects (dashed line) in one-year bins, with RWB source fraction indicated. There is a trend of increasing RWB fraction with longer time separation. 
\label{fig:mjd}}
\end{figure}

While most sources show BWB trend, the RWB trend is clearly present in our sample, although very rare (see Figs. 2 and 5). In our sample, DR7 epochs usually span about four years, while the time coverage extends to about ten years when adding BOSS spectra. We found a positive correlation between $\alpha_{\rm d}$ and time separation $\Delta \rm MJD$ of bright and faint epochs with a Spearman correlation coefficient $r_{\rm s} = 0.10$ at 99.99$\%$ confidence level (see Fig. \ref{fig:mjd}). Moreover, we found a trend of increasing RWB source fraction at longer time separation. The results imply that variable emission at longer timescales possibly have a higher probabilities to show RWB trend than that at short timescales. The result is consistent with previous work based on 9258 photometric quasars in Stripe 82 \citep{sun2014}, which is expected by the time-dependent inhomogeneous disk model \citep{dexter2011}.

Assuming that variability occurs in the inner region of AGNs, so when the accretion disk becomes hotter, it will produce more high energy photons and the continuum emission peak will move to short wavelength, yielding bluer spectra when AGNs become brighter \citep{bian2012}. While this can generally explain BWB trend, the reason of RWB trend is still unclear. Based on the optical variability study of eight red blazars, \cite{gu2006} suggested that the different relative contributions of the thermal versus non-thermal radiation in optical emission can be responsible for the observed color variations. The RWB trend observed in flat-spectrum radio quasars (FSRQs) can be explained by the significant contribution of thermal emission in optical/UV band, especially at low flux state. In FSRQs, the flux variations are usually dominated by the variable jet emission. When the source gets brighter, the jet emission will become gradually dominant, resulting in a RWB trend since jet emission is redder than thermal emission from accretion disk. However, we found that radio-loud quasars in our sample, especially objects with R $\geq$ 100, do not show significant higher probability of RWB trend (see Fig. \ref{fig:comp}), indicating that the non-thermal jet emission may not be a dominant factor in producing RWB trend in present sample. On the other hand, since luminous quasars are usually hosted in bright elliptical galaxies, the host galaxies will be more extended in poor seeing conditions, then contribution of host galaxies will be smaller in a fixed aperture. This will lead a RWB trend when seeing varies because the host galaxies are usually redder than quasars \cite[see e.g., ][]{guo2014}. In order to check seeing effect, we preformed correlation analysis between the spectral index and seeing for a subsample of RWB quasars at redshift of 0.3 to 0.8. However, no significant correlation was found from Spearman correlation analysis ($r_{\rm s} = - ~0.04$ at $\sim 76\%$ confidence level). This shows that RWB trend is likely not much related with seeing variation. Actually, this could be a natural result of our sample selection. Our sample is selected at redshift of $z>0.3$ in order to reduce the contamination of host galaxies. At such redshift, the contribution of host galaxies is expected to be rather small, and it unlikely changes much when seeing varies.

In accretion disk, the optical/UV radiation originates from inner hot region, and the light is bluer than that from outer cool region. The radiation regions may play an important role in color variation as shown in \cite{shields1978}. Regardless of the detailed mechanism of variability (e.g., localized thermal fluctuations or large temperature fluctuations), when the variation occurs firstly in outer disk region, it will take certain time to propagate the fluctuation inward. The RWB trend will happen when inner disk region has not fully responded yet. On the contrary, if fluctuations happen firstly in inner region, the BWB trend will be observed. Since most of our sources have BWB trend, it's likely that the fluctuations are more often happening firstly in inner region, while rarely occurs in outer disk, although the details are unknown. Obviously, the further investigation on the color variations on a much larger sample will be needed to improve our understanding on variability, such as, using the massive data from Large Synoptic Survey Telescope (LSST).

\subsection{Compare with previous work}
While the result of BWB in most of our sources is consistent with SDSS-based works \cite[e.g.,][]{ruan2014}, it is different from \cite{bian2012}, in which half of the quasars appear redder during their brighter phases, not only for variable radio-quiet quasars, but also for variable radio-loud objects. The difference may be related with several factors. Firstly, their sample is from the FIRST Bright Quasar Survey \cite[FBQS,][]{whi00}, therefore basically is a radio-selected sample. All their sources were detected in FIRST, of which there are 312 radio-loud and 232 radio-quiet quasars. However, our sample is optically selected, with only 246 objects included in FIRST catalog. The jet contribution in continuum emission has been proposed to explain the RWB found in blazars \cite[e.g.,][]{gu2006}. The presence of a radio jet thus likely affect the color variations towards RWB to some extent, although not completely. Indeed, the RWB source fraction is about 11\% in our 246 FIRST-detected quasars, higher than that in whole sample ($\sim6\%$). Secondly, a significant number of sources in their sample are at low redshift, with about one third of quasars at z $\le$ 0.5. The original spectra of FBQs used to compare with SDSS spectra in their work, are from observations at five different observatories, in a wide variety of atmospheric conditions, ranging from photometric to cloudy, with both good and bad seeing, and different resolutions \citep{whi00}. The variations of host galaxy contribution when seeing varies were proposed to be at least partly responsible for RWB trend \citep{guo2014}. This effect could be more significant at low redshift, especially when comparing the spectra taken with long slit and fiber. While our analysis shows that RWB trend is likely not much related with seeing variation (see Section 5.5), it is unclear in their sample. Finally, while we found a trend of increasing fraction of RWB with longer spectral duration (Fig. \ref{fig:mjd}), their sample more concentrates on larger time separations (i.e., > 3 yrs, see their Fig. 1). This could be partly the reason for the high RWB fraction.

\section{Conclusions}

We investigated the optical/UV color variations for a sample of 2169 quasars based on the multi-epoch spectroscopy in DR7 and DR9 by correcting their systematic difference. The color variation was explored by comparing the difference of power-law spectral
index of the continuum between the brightest and the faintest spectra, and by the shape of the difference spectrum. In 1876 quasars with consistent color variations from two methods, we found that most sources (1755, $\sim 94\%$) show BWB trend, and the RWB trend is only detected in 121 objects ($\sim 6\%$). The color variations are generally uniform and independent of  black hole mass, redshift, Eddington ratio and luminosity. The common BWB trend is supported by the bluer composite spectrum constructed from bright spectra than that from faint spectra, and also by the blue composite difference spectrum. The correction spectrum is proved to be highly reliable by comparing the composite spectrum from corrected DR9 and original DR7 spectra. It thus can be widely applied in future work to correct the systematic difference between SDSS-DR7 and BOSS.  Assuming that the optical/UV variability is triggered by fluctuations, RWB trend can likely be explained if fluctuations occur firstly at outer disk region, and the inner disk region has not fully responded yet when fluctuations being propagated inward. The common BWB trend implies that fluctuations are likely more often happening firstly in inner disk region.

We especially thank the anonymous referee for his/her thorough report and helpful comments and suggestions that have significantly improved the paper. We thank Shuang-liang Li and Junxian Wang for valuable discussions. This work is supported by the National Science Foundation of China (grants 11473054, 11373056, and U1531245) and by the Science and Technology Commission of Shanghai Municipality (14ZR1447100). This work makes extensive use of SDSS-I/II and SDSS-III data. The SDSS-I/II Web Site is http://www.sdss.org/. The SDSS-III Web Site is http://www.sdss3.org/.

{}

\end{CJK}
\end{document}